\documentclass[preprint2]{aastex}
\usepackage{epsf}

\slugcomment{For submission to ApJL.}

\shorttitle{Moment equations and numerical simulations}
\shortauthors{F. Kupka}

\begin{document}

\title{Turbulent convection: comparing the moment equations \\
       to numerical simulations.}

\author{F. Kupka\altaffilmark{1}}
\affil{Institute for Astronomy, University of Vienna,
       T\"urkenschanzstra{\ss}e 17, A-1180 Wien, Austria}
\email{kupka@astro.univie.ac.at}

\altaffiltext{1}{Institute of Mathematics, University of Vienna,
                 Strudlhofgaase 4, A-1090 Wien, Austria}

\begin{abstract}
The non-local hydrodynamic moment equations for compressible convection 
are compared to numerical simulations. Convective and radiative flux typically 
deviate less than 20\% from the 3D simulations, while mean thermodynamic 
quantities are accurate to at least 2\% for the cases we have 
investigated. The moment equations are solved in minutes rather than days 
on standard workstations. We conclude that this convection model has the 
potential to considerably improve the modelling of convection 
zones in stellar envelopes and cores, in particular of A and F stars. 
\end{abstract}


\section{Introduction}

In a wide range of astrophysical problems heat is transported by both 
radiation and convection. Examples include stellar envelopes and stellar 
cores, where convection may be coupled with rotation, pulsation, 
and diffusion. To be useful, a convection  model must be both manageable 
and reliable. Local convection models satisfy both criteria but they are 
restricted to regions of strong to moderately strong convection, while 
in most cases one needs to model moderately strong to weak convection. 

Numerical simulations (\citet{Sofia84,Freytag96}, and \citet{Atro94,Kim98,
Stein98}) have been applied to 2D and 3D stellar atmospheres. Their successful 
reproduction of spectral line profiles and solar granulation statistics has 
proved their reliability. However, they are restricted to layers near the 
stellar surface \citep{Kup99a} because of the huge thermal time scales 
characterizing stellar interiors. In addition, they currently are too 
expensive for applications such as non-linear stellar pulsation calculations 
for RR Lyrae stars \citep{Feu98} or for use in everyday spectrum synthesis of 
large wavelength ranges or large parameter sets for A--M stars. Hence, they 
do not satisfy the criterion of manageability for the whole range 
of problems where convection occurs.

Convection models based on the non-local, hydrodynamic moment equations 
provide a possible alternative. They describe the moments of an ensemble 
average of the basic fields: velocity $U$, temperature $T$, and density $\rho$ 
(or pressure $P$). Dynamic equations for the moments are derived directly 
from the fully compressible Navier-Stokes equations (NSE, \citet{Can97}). 
The  ``ensemble'' used in this averaging process consists of realizations of 
solutions of the NSE \citep{Xiong86,Xiong97,Can92,Can93,Can97,CD98}, or of 
``convective elements'' of different velocity and temperature 
\citep{Gross93,Gross96}. However, the moment equations entail higher order 
moments and thus require closure assumptions. To obtain a closed set of 
equations most convection models have invoked a mixing length. The models 
derived in Canuto (1992, 1993, 1997) and in \citet{CD98} avoid a mixing 
length by providing a fully non-local set of dynamic equations for the second 
order moments. 

Recently, \citet{Kup99a} has presented the first comparison of a variant of
these convection models with numerical simulations of compressible convection 
for a stellar-like scenario. He used the downgradient approximation (DGA) for 
the third order moments. To avoid its shortcomings, here we use a more 
complete model. It is based on a model introduced in \citet{Can92}. The latter
was successfully applied to the convective boundary layer of the terrestrial 
atmosphere \citep{Can94}. We describe the physical scenario for our comparison 
and present results for two sample problems. We also consider the potential of 
such models for application to envelope convection in A and F stars.

\section{The physical scenario}

In \citet{Muth95a,Muth99a} the fully compressible NSE were solved for
a 3D plane parallel geometry with a constant 
gravity $g$ pointing to the bottom of a simulation box. Periodic boundary 
conditions were assumed horizontally. A constant temperature $T$ was 
prescribed at the top and a constant input flux at the bottom of the box
($\partial T/\partial z = {\rm const.}$, $z$ denotes the vertical 
direction, i.e.\ top to bottom). Top and bottom were taken to be
impenetrable and stress free. A perfect gas law was assumed and
radiation was treated in the diffusion approximation. Stable and unstable 
layers were defined through $\partial T/\partial z$, which initially was
a piecewise linear function in units of the adiabatic gradient: 
$\partial T/\partial z = b(z) (\partial T/\partial z)_{\rm ad}$, where 
$b(z)=1+(\nabla-\nabla_{\rm ad})/\nabla_{\rm ad}$. To define the stability 
properties of the layers a Rayleigh number ${\rm Ra}$ for one zone where 
$b(z)={\mbox{\rm const.}}$ and $>1$ \citep{Muth95a,Muth99a} is specified
together with a Prandtl number ${\rm Pr}$ and an initial temperature contrast.
This yields the radiative conductivity $K_{\rm rad}(z)$. Both ${\rm Pr}$ and 
$K_{\rm rad}$ are kept fixed and place regions of stable and unstable 
stratification at different vertical locations in the simulation box. 
A similar approach was used by \citet{Catt91, Chan96, Hurl94, Port94, Singh95} 
and others. The comparison to simulations with a prescribed viscosity avoids 
the need to use a subgrid scale model. According to our numerical experiments
with the moment equations, molecular viscosity can decrease the efficiency of 
convection as measured by $F_{conv}/F_{total}$ by up to 15\% and smooths out 
the numerical solution in stably stratified regions \citep{Kup99a,Kup99b}.

\section{A convection model based on the hydrodynamic moment equations}

Using the Reynolds stress approach, \citet{Can92,Can93} has derived 
a convection model which consists of four differential equations for
the basic second order moments $K$ (turbulent kinetic energy),
$\overline{\theta^2}$ (mean square of temperature fluctuations, i.e.\ thermal 
potential energy), $\overline{w\theta}$ ($F_{\rm c} = c_p \rho J = c_p \rho 
\overline{w\theta}$ is the convective flux), and 
the vertical turbulent kinetic energy $\overline{w^2}$. The model was 
rederived by \citet[CD98]{CD98} using a new turbulence model based on 
renormalization group techniques. In CD98 notation, the convection model 
reads ($\partial_t \equiv \partial/\partial t$, 
$\partial_z \equiv \partial/\partial z$):
\begin{equation}
  \partial_t K + D_{\rm f}(K) = g \alpha J - \epsilon
        + \frac{1}{2}C_{ii} + \partial_z (\nu \partial_z K), 
\end{equation}
\begin{eqnarray}
 \lefteqn{\partial_t(\frac{1}{2}\overline{\theta^2})
      + D_{\rm f}(\frac{1}{2}\overline{\theta^2})  =   \beta J
        - \tau_{\theta}^{-1}\overline{\theta^2} } \nonumber\\
        & & {} + \frac{1}{2}\partial_z (\chi
          \partial_z \overline{\theta^2}) + \frac{1}{2}C_{\theta},
\end{eqnarray}
\begin{eqnarray}
 \lefteqn{\partial_t J + D_{\rm f}(J)  =  \beta \overline{w^2}
        + \frac{2}{3} g \alpha \overline{\theta^2}  
        - \tau_{{\rm p}\theta}^{-1} J  } \nonumber\\
        & & {} + \frac{1}{2}\partial_z (\chi \partial_z J)    
        + C_3 + \frac{1}{2}\partial_z (\nu \partial_z J),
\end{eqnarray}
\begin{eqnarray}
 \lefteqn{\partial_t(\frac{1}{2}\overline{w^2}) + D_{\rm f}(\frac{1}{2}
        \overline{w^2}) 
        = - \tau_{\rm pv}^{-1}(\overline{w^2}-\frac{2}{3}K) } \nonumber\\
        & & {} + \frac{2}{3} g \alpha J - \frac{1}{3}\epsilon 
        + \frac{1}{2}C_{33} + \frac{1}{2}\partial_z (\nu
                 \partial_z \overline{w^2})
\end{eqnarray}
\begin{eqnarray}  \label{eq_epsilon}
 \lefteqn{ \partial_t \epsilon + D_{\rm f}(\epsilon) =
     c_1 \epsilon K^{-1} g \alpha J - c_2 \epsilon^2 K^{-1}
       + c_3 \epsilon \tilde{N} } \nonumber\\
     & & {} + \partial_z (\nu\partial_z \epsilon),
         \quad \tilde{N} \equiv \sqrt{g \alpha |\beta|},
\end{eqnarray}
\begin{equation}    \label{eq_diffeps}
   D_{\rm f}(\epsilon) \equiv \partial_z(\overline{\epsilon w})
   = -\frac{1}{2}\partial_z
   \left[(\nu_{\rm t} + \chi_{\rm t})\partial_z \epsilon \right]
\end{equation}
Here, $\alpha$ is the volume compressibility ($=1/T$ for a perfect gas),
$\beta$ is the superadiabatic gradient, $\chi = K_{\rm rad}/{c_p \rho}$, and
$\nu = \chi {\rm Pr}$. The $\tau$'s are time scales. We use (25a), (27b), and 
(28b) of CD98 to relate the latter to the dissipation time scale $\tau = 2 K / 
\epsilon$. Compressibility effects are represented by $C_{ii}, C_{\theta}, C_3, 
C_{33}$ given by equations (42)--(48) of \citet{Can93}. We neglected a few
terms of $C_{ii}, C_{\theta}, C_3, C_{33}$ that were too small to contribute 
to the solution of the moment equations \citep{Kup99b}. 
In equation (\ref{eq_diffeps}), $\nu_{\rm t} = C_{\mu} 
K^2/\epsilon$, and $C_{\mu}$ is a constant given by (24d) of CD98, for which 
we take the Kolmogorov constant ${\rm Ko} = 1.70$, while $\chi_{\rm t}$ is 
given by the low viscosity limit of (11f) of CD98. We optionally 
included molecular dissipation by restoring the largest (i.e.\ second order 
moment) terms containing the kinematic viscosity $\nu$ (i.e.\ $\partial_z 
(\nu \partial_z K)$, etc.). They are important when ${\rm Pr}$ is of order 
unity rather than zero \citep{Kup99a,Kup99b}. Hence, we 
included them in all the examples shown below. For the same reason, we 
optionally included a term in (\ref{eq_epsilon}) that accounts for molecular 
dissipation effects. We use $c_1=1.44$, $c_2=1.92$, and $c_3=0.3$ where 
$\beta < 0$ while $c_3=0$ elsewhere (as in CD98). The local limit of 
(\ref{eq_epsilon}), $\epsilon = K^{3/2} / \Lambda$ with $\Lambda = 
\alpha H_{\rm p}$ and $H_{\rm p} = P / (\rho g)$, fails to yield reasonable 
filling factors \citep{Kup99a} and was thus avoided.  

To calculate the mean stratification we solve
\begin{eqnarray}  
   \partial_z (P + p_{\rm t})  & = & - g \rho, 
        \label{eq_hydrostat} \\
   c_{\rm v}\rho \partial_t T +
            \rho \partial_t K & = & -\partial_z
            (F_{\rm r} + F_{\rm c} + F_{\rm k})  \label{eq_Temp}
\end{eqnarray}
where $p_{\rm t}  =  \rho \overline{w^2}$. Equations
(\ref{eq_hydrostat})--(\ref{eq_Temp}) are taken from equation (103)
of \citet{Can93} (excluding the higher order term in his equation (104)), 
and from equation (18c) of CD98 (for the latter, $c_p$ was
substituted to $c_v$ to account for non-Boussinesq effects, Canuto, priv.\ 
communication). In the stationary limit, (\ref{eq_Temp}) yields 
$F_{\rm total} = F_{\rm r} + F_{\rm c} + F_{\rm k}$
where the kinetic energy flux is given by $F_{\rm k}=\rho\overline{Kw}$
and the radiative flux $F_{\rm r} = -K_{\rm rad} \partial T/\partial z$.

Non-locality is represented by the terms $D_{\rm f}(K)$, 
$D_{\rm f}(\frac{1}{2}\overline{\theta^2})$, $D_{\rm f}(J)$, and 
$D_{\rm f}(\frac{1}{2}\overline{w^2})$, which require third order moments 
(TOMs). Using the downgradient approach (DGA) for the TOMs \citet{Kup99a} 
found qualitatively acceptable results for the convective flux and filling 
factors. But quantitatively the results were not satisfactory,
which corroborated the shortcomings of the DGA found in \citet{Can94} and in 
\citet{Chan96}. To improve over the DGA, one 
must solve the dynamic equations for the TOMs (see \citet{Can92,Can93}). 
We investigated both the fully time dependent case which requires 6 additional 
differential equations and various approximations of their stationary limit 
which do not entail further differential equations. Here, we consider the 
following ``intermediate'' model for the TOMs: we take the stationary limit of 
the dynamic equations given in \citet{Can93}, but neglect the Boussinesq 
terms which depend on $\beta$ \citep{Kup99b}. This model is similar in 
robustness to the DGA, but yields significantly better results.

For the boundary conditions we impose 
$T = {\rm const.}, \overline{w^2} = 0, J = 0, \partial K/\partial z = 0, 
\partial\epsilon/\partial z = 0, \overline{\theta^2} = 0$ at the top, while 
at the bottom $\partial T/\partial z = 0, \overline{w^2} = 0, J = 0,
\partial K/\partial z = 0, \partial\epsilon/\partial z = 0, \partial 
\overline{\theta^2}/\partial z = 0$. This choice permits stable numerical 
solutions which are consistent with the boundary conditions for the 
numerical simulations. The equation for the mean pressure is constrained by 
varying $P$ at the top such that for a given $T$ the resulting density 
stratification is consistent with mass conservation.

\section{Results and discussion}

We have used simulation data representing two configurations: model 3J, 
a convection zone embedded between two strongly stable layers \citep{Muth95a}, 
and model 211P, a stably stratified layer embedded between two unstable layers
(with a small stable layer at the bottom, \citet{Muth99a}). In both cases, 
radiation transports at least 80\% of the input energy and ${\rm Pr} = 1$.
Each unstable layer has a thickness of $\approx 0.5$--$2.5$ $H_{\rm p}$. 
Model 3J has an initial adiabatic temperature contrast of
3.5 and encompasses 4.2 $H_{\rm p}$ while model 211P has a contrast of 6.0 and 
encompasses about 4.8 $H_{\rm p}$. The numerical simulations were done for 
$72 \times 50 \times 50$ grid points and have successfully been compared to 
higher resolution simulations ($125 \times 100 \times 100$ grid points, 
\citet{Kup99c}). All simulation data shown here are statistical averages 
over many dozen sound crossing times.

The moment equations were solved by centered finite differences on a staggered 
mesh. Solutions were calculated using 72 grid points and were successfully 
verified by comparing them with results obtained from higher resolutions of 
128 to 512 grid points. Time integration was done by the Euler forward method 
until a stationary state was reached after 1.2 to 2.5 thermal time scales. 
Verification of stationarity was done by testing whether all errors in time 
derivatives were formally less than $10^{-8}$ in relative units at all 
grid points (i.e.\ far smaller than the truncation error) and by checking 
the strict energy flux and mass conservation required for stationary 
solutions.

Fig.~\ref{Fig1} compares the convective flux of model 211P with two convection 
models: the non-local model with intermediate TOM and its local counterpart 
(the stationary, local limit of the non-local model, see CD98). Clearly, only 
the non-local model successfully reproduces the simulations also in the central 
overshooting region which connects both convection zones of model 211P. 

Fig.~\ref{Fig2} compares the filling factor $\sigma$, i.e.\ the relative area 
in each layer covered by upwards flowing material, of the numerical
simulations of model 3J with the filling factor computed from the
non-local moment equations. For the latter we used the prescription described
in CD98. As $\sigma$ is an important topological quantity describing the 
inhomogenous nature of convection, the reasonable agreement found here is 
a very promising result. It illustrates the importance of improving the TOMs, 
because the DGA used in \citet{Kup99a} only indicated a correct trend, while 
the more complete TOM used here also provides a closer quantitative agreement.
The local convection model predicts a structureless $\sigma=0.5$.

Fig.~\ref{Fig3} compares the convective flux from numerical simulations for 
model 3J with solutions from the moment equations using the intermediate TOM. 
Clearly, the latter has improved over the DGA \citep{Kup99a} both qualitatively 
and quantitatively. Except for the overshooting region the DGA barely
differed from the local model which is shown here for comparison. Though 
the present convection model provides a substantial improvement,
the convective flux still falls short in the middle of the convection zone of 
3J and the extent of the lower overshooting zone is underestimated. This may 
be due to neglecting effects of radiative losses on the time scales 
$\tau_{\theta}$ and $\tau_{p\theta}$ as well as because several contributions 
to the TOMs are neglected in our ``intermediate model'', or due to the 
Boussinesq treatment of the TOMs in \citep{Can92,Can93}, or incompleteness of 
equation (\ref{eq_epsilon}).

Fig.~\ref{Fig4} compares local length scales $\Lambda =
\alpha H_{\rm p}$ as used in mixing length theory with the length scale 
$\Lambda = K^{1.5}/\epsilon$ obtained with the non-local moment equations
for the case of model 211P. Obviously, there is no $\alpha$ that brings the 
length scales in agreement. Thus, there is no alternative but to solve at
least the full equations (\ref{eq_epsilon})--(\ref{eq_diffeps}) to obtain
$\epsilon$.

In conclusion, we have found not only qualitative, but also quantitative 
agreement between numerical simulations and the non-local moment equations, 
provided one avoids the DGA for the TOMs and employs the stationary solution 
of their dynamic equations as suggested in appendix B of \citep{Can93} and 
neglects Boussinesq type factors (involving $\beta$). Moreover, we have 
found the mean values for T, P, and $\rho$ to be accurate to within
2\% in comparison with 4\% found for local models with optimized mixing 
length parameter. Convective and radiative flux are typically accurate to 
within 20\%. The improvements are largest in regions of weak convection and 
in stably stratified layers. We have used the closure constants suggested in 
\citet{Can93} and CD98 except for $c_{11}$, which was increased from 0.2 to 
0.5, because this improved the results for the planetary boundary layer 
(Canuto priv.\ communication) and enhanced the numerical 
stability. We have not found the DGA to be able to yield a similar agreement
for both 3J and 211P even when tuning the closure constants individually
for 3J and 211P. The situation is even worse for the local model. In
\citet{Kup99b} we will deal with this problem in detail. Finally, while
each 3D simulation took between several days and several 
weeks on a modern workstation to obtain a thermally relaxed solution, the 
moment equations took a couple of minutes to half an hour. This holds true 
already for a low numerical resolution and for an explicit time 
integration method. 

The results found here are promising for the application of the non-local 
convection model in particular to envelope convection zones of A and F stars, 
because they feature a similar range of convective efficiency, thickness in 
terms of $H_p$, and interaction among neighbouring convection zones. For 
A and F stars the thermal structure is not known in advance. Hence, thermal 
relaxation and thus the gain on speed in comparison with numerical simulations 
is essential. The computational savings are very attractive also for problems
studied on hydrostatic time scales, whenever the full information provided 
by a simulation is not needed. Improvements of the convection model studied 
are possible, nevertheless it is a more promising basis for 
asteroseismological studies of pulsating A and F stars ($\delta$ Sct, 
$\gamma$ Dor, roAp, etc.) than local convection models and also a new basis 
for related work such as diffusion calculations. Detailed results for 
a broader range of physical parameters, in particular for efficient convection 
and deeper convection zones, and thus of high importance also for other types 
of stars, have yet to corroborate this study. 

\acknowledgments
I am indebted to H.J.~Muthsam for permission to use his simulation
code and data. I am grateful to V.M.~Canuto and M.S.~Dubovikov 
for discussions on turbulent convection models.
The research was performed within project {\sl P11882-PHY} ``Convection 
in Stars'' of the Austrian Fonds zur F\"orderung der wissen\-schaft\-lichen 
Forschung.

\clearpage

\figcaption[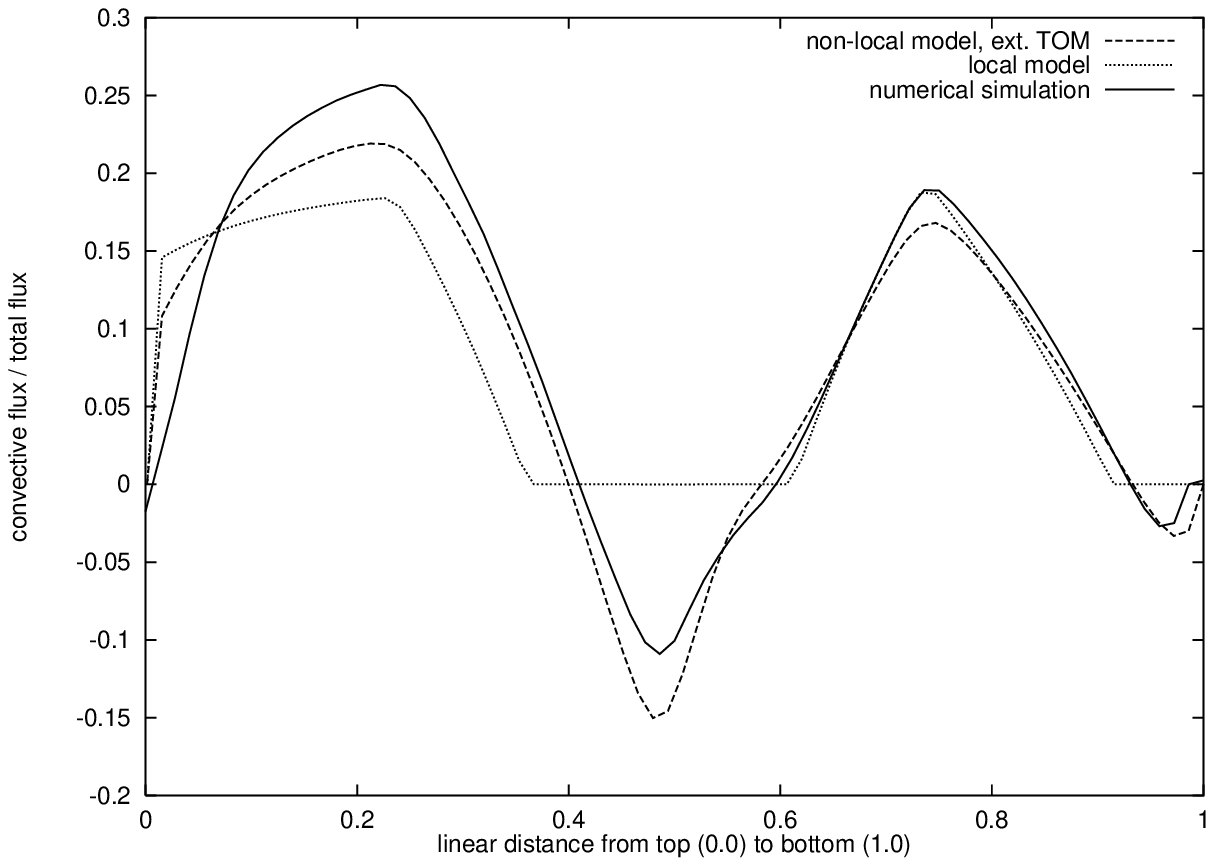]{Convective flux (model 211P) for the non-local convection 
model and for its local limit (mixing length
$\Lambda = H_{\rm p}$), and a numerical simulation. \label{Fig1}}

\figcaption[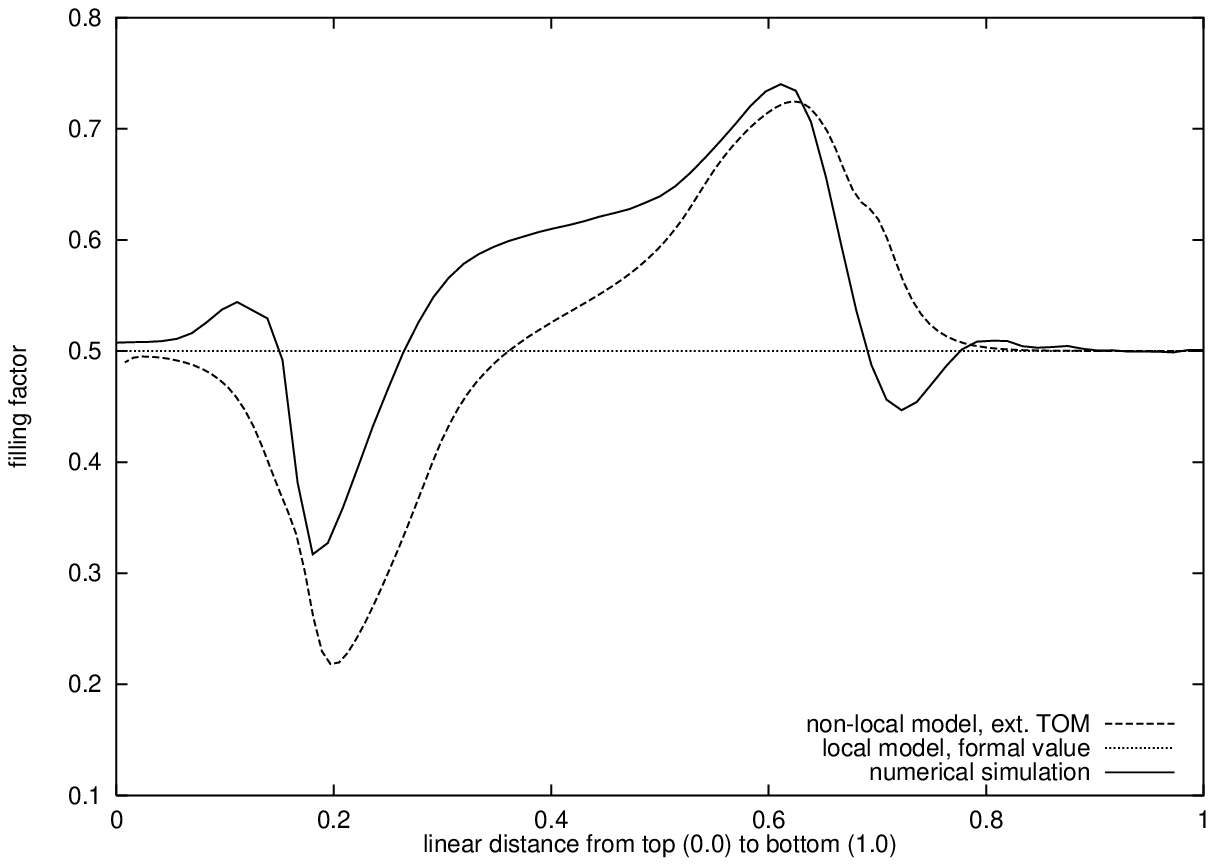]{Filling factor $\sigma$ (model 3J) for the non-local 
convection model and a numerical simulation. For the local convection model 
$\sigma = 0.5$. \label{Fig2}}

\figcaption[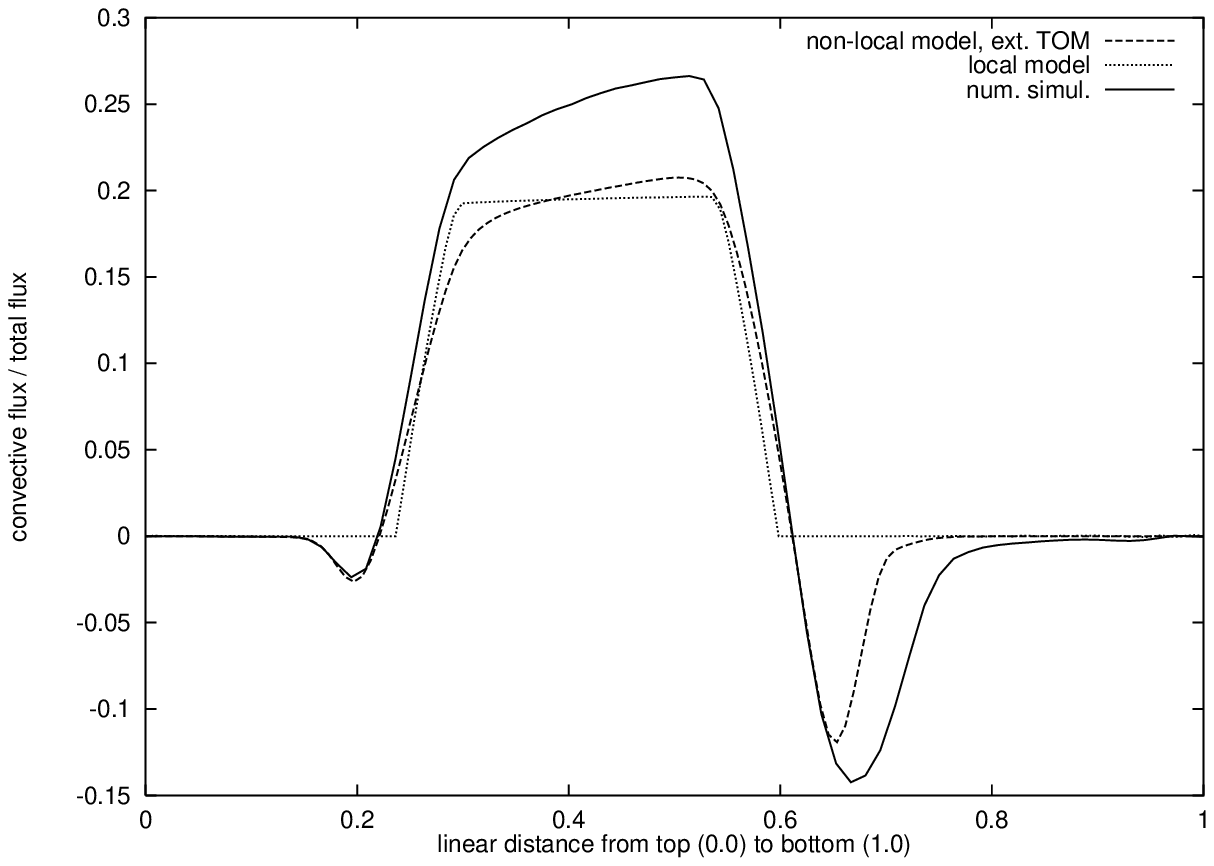]{Convective flux (model 3J) for the non-local convection 
model and for its local limit (mixing length
$\Lambda = H_{\rm p}$), and a numerical simulation. \label{Fig3}}

\figcaption[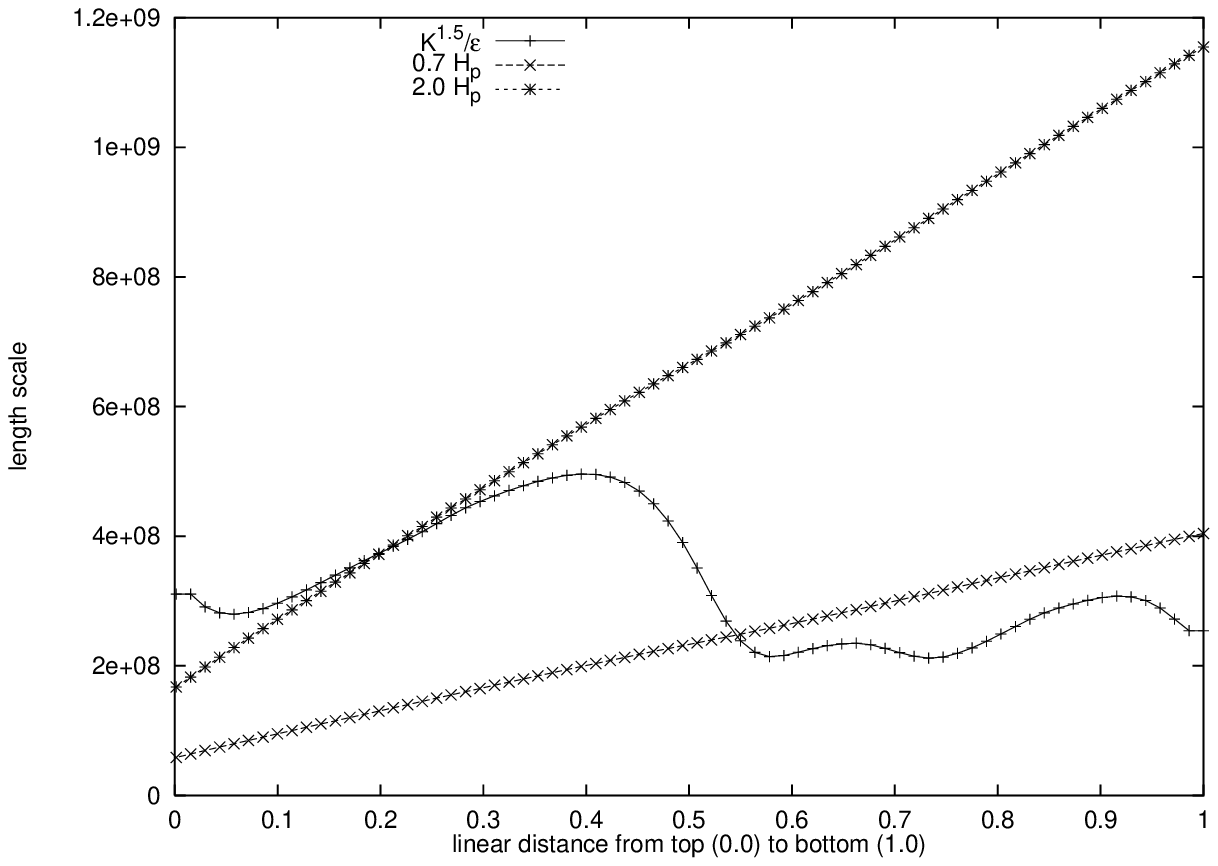]{Length scale $\Lambda = K^{1.5}/\epsilon$ (model 
211P) from the non-local convection model compared with $\Lambda=\alpha 
H_{\rm p}$ for $\alpha = 0.7, 2.0$.  \label{Fig4}}

\clearpage

\begin{figure}
\epsffile{fig1_big.eps}
\end{figure}

\clearpage

\begin{figure}
\epsffile{fig2_big.eps}
\end{figure}

\clearpage

\begin{figure}
\epsffile{fig3_big.eps}
\end{figure}

\clearpage

\begin{figure}
\epsffile{fig4_big.eps}
\end{figure}

\end{document}